\documentclass[
    ,final            
  ]
  {aipproc}

\layoutstyle{6x9}
\usepackage{epsf,graphics,amssymb,amsfonts,amsmath,array}

\def\bfnabla{\mbox{\boldmath $\nabla$}}

\def\bfsigma{\mbox{\boldmath $\sigma$}}

\def\em{{\rm em}}
\def\lQ{\Lambda_{\rm QCD}}

\def\als{\alpha_{\rm s}}

\def\siml{{\ \lower-1.2pt\vbox{\hbox{\rlap{$<$}\lower6pt\vbox{\hbox{$\sim$}}}}\ }} 
\def\simg{{\ \lower-1.2pt\vbox{\hbox{\rlap{$>$}\lower6pt\vbox{\hbox{$\sim$}}}}\ }}

\def\vbfD{{\ \lower-8pt\vbox{\hbox{\rlap{$\!\leftrightarrow$}\lower8pt\vbox{\hbox{$\!\bf D$}}}}\ }} 
\def\dsl{\,\raise.15ex\hbox{/}\mkern-13.5mu D}

\newcommand{\nn}{\nonumber}
\newcommand{\be}{\begin{equation}} 
\newcommand{\ee}{\end{equation}}
\newcommand{\bea}{\begin{eqnarray}} 
\newcommand{\eea}{\end{eqnarray}}
\newcommand{\beq}{\begin{equation}}
\newcommand{\eeq}{\end{equation}}
\newcommand{\bqa}{\begin{eqnarray}}
\newcommand{\eqa}{\end{eqnarray}}
\newcommand{\bra}[1]{\ensuremath{\langle#1|}}
\newcommand{\ket}[1]{\ensuremath{|#1\rangle}}

\begin{document}

\title{Heavy quarkonium decays and transitions in the language of effective
  field theories}


\author{Antonio Vairo}{
  address={Dipartimento di Fisica dell'Universit\`a di Milano and INFN, via
  Celoria 16, 20133 Milano, Italy}}

\begin{abstract}
Heavy quarkonium decays and transitions are discussed in the 
framework of non-relativistic effective field theories.
Emphasis is put on the matching procedure in the non-perturbative 
regime. Some exact results valid for the magnetic dipole couplings  
are discussed.
\end{abstract}

\maketitle

\section{Introduction}
In the last years the B factories, CLEO and BES have produced a large amount 
of new data for heavy quarkonium observables \cite{Brambilla:2004wf}.
These data are not only interesting because they may signal 
new states or new decay or production mechanisms, but also because 
heavy quarkonium is a system that to a large extent can rigorously be  
studied in QCD. Therefore, any new understanding of it may potentially 
provide new insight on the non-perturbative dynamics of QCD.

Heavy quarkonium, as a non-relativistic bound state, is characterized by a hierarchy of 
energy scales: $m$, $mv$ and $mv^2$, where $m$ is the heavy-quark mass 
and $v\ll 1$ the heavy-quark relative velocity. 
Whenever a system is described by a hierarchy of scales, observables 
may be calculated by expanding one scale with respect to the other. 
An effective field theory (EFT) is a field theory that makes this expansion 
explicit at the Lagrangian level.
To be more precise, let us call $H$ a system described by a fundamental
Lagrangian $\cal L$ and suppose it characterized by 2 scales: $\Lambda \gg \lambda$. 
The EFT Lagrangian,  ${\cal L}_{\rm EFT}$ , suitable to describe  $H$ 
at scales lower than $\Lambda$, is characterized by (1) a cut off  $\Lambda \gg \mu \gg \lambda$; 
(2) some  degrees of freedom  that exist at scales lower than $\mu$.
The Lagrangian ${\cal L}_{\rm EFT}$ is then made of all operators  $O_n$ 
that may be built from the effective  degrees  
of freedom  and are consistent with the  symmetries of the original Lagrangian
$ {\cal L}$:
\be
{\cal L}_{\rm EFT}  = \sum_n c_n(\Lambda/\mu)   \frac{O_n(\mu,\lambda)}{\Lambda^n}.
\label{EFT}
\ee
The advantage is that, once the scale $\mu$ has been run down to $\lambda$, 
the power counting is homogeneous $\langle  O_n  \rangle
\sim  \lambda^n$, so that  the EFT is, indeed, organized as an expansion in  $\lambda/\Lambda$. 
Despite the EFT not being renormalizable in the traditional sense, it is 
renormalizable order by order in $\lambda/\Lambda$. 
The  matching coefficients  $c_n(\Lambda/\mu)$  encode the non-analytic
behaviour in  $\Lambda$. They are calculated by imposing 
that   ${\cal L}_{\rm EFT}$  and   ${\cal L}$  describe the same
physics at any finite order in the expansion. The procedure is called  matching.
Finally, we note that if $ \Lambda \gg \lQ$, $c_n(\Lambda/\mu)$ may be
calculated in perturbation theory, if $ \Lambda \sim \lQ$, the matching 
must rely on non-perturbative methods.

Several effective field theories for heavy quarkonium that take full advantage 
of the non-relativistic hierarchy of scales have been developed and used 
over the last decade. For a recent review we refer to \cite{Brambilla:2004jw}.
NRQCD is the EFT that exploits the hierarchy  $\Lambda =m \gg \lambda = mv$
\cite{Caswell:1985ui,Bodwin:1994jh}. Since $m \gg \lQ$, the matching
coefficients of NRQCD may be calculated in perturbation theory.
pNRQCD is the EFT that exploits the hierarchy  $\Lambda =mv \gg \lambda = mv^2$
\cite{Pineda:1997bj,Brambilla:1999xf}.  If $\lQ \sim mv^2$, then the matching
to pNRQCD may be still done in perturbation theory. We call weak coupling
this regime, which may be suited to describe ground-state quarkonium. 
If $\lQ \sim mv$, then the matching to pNRQCD is non
perturbative. We call strong coupling this regime, which may be suited to 
describe excited quarkonium states. 

The fact that EFTs may be built to describe heavy quarkonium 
in the strong-coupling regime follows from the observation  
that the non-relativistic hierarchy of scales survives also below 
$\lQ$ \cite{Brambilla:2000gk,Pineda:2000sz}.
The complication of the strong-coupling regime comes from the non-perturbative
matching and from new scales that may arise in loops sensitive to $\lQ$.
An example is the three-momentum scale $\sqrt{m\lQ}$ 
discussed in \cite{Brambilla:2003mu}. Nevertheless, many advantages 
remain in treating even strongly-coupled heavy quarkonium in an EFT framework. 

In the following we shall sketch a unified framework for the description 
of inclusive and electromagnetic decays, and radiative transitions in the framework of
strongly coupled pNRQCD. For a treatment of inclusive and electromagnetic
decay widths in the weak-coupling regime we refer to \cite{Brambilla:2004jw} 
and references therein. For a treatment of magnetic dipole transitions 
in the weak-coupling regime we refer to \cite{BJV05}.

\section{NRQCD}
NRQCD is the EFT that follows from QCD when modes of energy or momentum 
$m$ are integrated out. The structure of the EFT Lagrangian is like  
Eq.~(\ref{EFT}) with $\Lambda = m$ and $\lambda = mv \sim \lQ$. 
The scale $mv$ is sometimes called soft. The degrees of freedom of the EFT Lagrangian are quarks, 
antiquarks and gluons with energy and momentum lower than $m$ (we neglect light quarks). 

The NRQCD Lagrangian may be written as 
\bea
{\cal L}_{\rm NRQCD} &=& {\cal L}_{2-f} + {\cal L}_{4-f} 
+ {\cal L}_{\rm light} \,,
\label{NRQCD:Lag}
\eea
where 
\bea
{\cal L}_{2-f} &=& 
\psi^\dagger \left( i D_0 + \frac{{\bf D}^2}{2m} \right) \psi
+ \frac{c_F}{2 m} \psi^\dagger \bfsigma \cdot g {\bf B} \psi
- \frac{2c_F -1}{8 m^2} \psi^\dagger \bfsigma \cdot [-i {\bf D}\times, g{\bf E}] \psi  
+ \dots \nn\\
&& + [\psi \rightarrow i \sigma ^2 \chi^*],
\label{LNRQCD2f}
\eea
\bea
{\cal L}_{\rm light} &=&
-\frac{1}{4}F^{\mu\nu\,a}F_{\mu\nu}^a .
\label{Llight}
\eea
$\psi$ is the Pauli spinor field that annihilates a heavy quark of mass $m$, 
$\chi$ is the corresponding one that creates a heavy antiquark,  
$i D_0 = i \partial_0 - g T^a A^a_0$  and  $i {\bf D} = i \bfnabla + g T^a {\bf A}^a$. 
The term ${\cal L}_{4-f}$ stands for the 4-fermion part of the NRQCD
Lagrangian (for an explicit expression see, for instance, \cite{Bodwin:1994jh}).

The coefficient $c_F$ is a matching coefficient
of the EFT. In Eq.~(\ref{LNRQCD2f}), we have made use of reparameterization invariance  
to reduce the other matching coefficients to this one. It is known at two loops
and may be found in \cite{Czarnecki:1997dz}.
The 4-fermion matching coefficients encode the contribution of the annihilation
graphs. As a consequence they develop an imaginary part. 
We refer to \cite{Vairo:2003gh} for an updated list of them and for references.

Let us give some definitions concerning the Fock space of NRQCD. 
If  $H_{\rm NRQCD}$ is the Hamiltonian of NRQCD, we call 
$|\underbar{n}; {\bf r} ,{\bf R}\rangle $ the
eigenstates of $H_{\rm NRQCD}$, and $E_n$ the corresponding eigenvalues. 
${\bf r}$ stands for the relative distance of the two heavy quarks 
and ${\bf R}$ for their centre-of-mass coordinate. Both are 
good quantum numbers in the static limit.
With $n$ we indicate a generic set of conserved quantum numbers. 
$|\underbar{n}; {\bf r} ,{\bf R}\rangle $ and 
$E_n({\bf r},{\bf R}; {\bfnabla}_{r}, {\bfnabla}_{R})$ satisfy the system of equations:
\bea 
&& \hspace{-8mm}
H_{\rm NRQCD} |\underbar{n}; {\bf r} ,{\bf R}\rangle = \int d^3r^\prime d^3R^\prime 
|\underbar{n}; {\bf r}^\prime ,{\bf R}^\prime \rangle 
E_n({\bf r}^\prime,{\bf R}^\prime; {\bfnabla}_{r^\prime}, {\bfnabla}_{R^\prime})
\delta^{3}({\bf r}^\prime-{\bf r})\delta^{3}({\bf R}^\prime-{\bf R}),
\label{bornschroe}
\\
&&\hspace{-8mm}
\langle \underbar{m}; {\bf r} ,{\bf R}|\underbar{n}; {\bf r}' ,{\bf R}'\rangle = 
\delta_{nm} \delta^{3} ({\bf r} -{\bf r}')\delta^{3} ({\bf R} -{\bf R}').
\label{normstate}
\eea

\section{Inclusive Decays in pNRQCD}
pNRQCD is the EFT that follows from NRQCD when gluons of energy or momentum 
and quarks of energy larger than $mv^2$ and quarks of momentum larger than $mv$ are integrated out. 
The structure of the EFT Lagrangian is like 
Eq.~(\ref{EFT}) with $\Lambda = mv \sim \lQ$ and $\lambda = mv^2$. 
The scale $mv^2$ is sometimes called ultrasoft. 
In the strong-coupling regime, if the gluonic excitations between the two heavy
quarks develop a mass gap of order $\lQ$, then they are all integrated out from
the theory. Therefore, the degrees of freedom of the EFT Lagrangian are 
only singlet quarkonium fields. The Lagrangian of pNRQCD is very simple:
\bea
{\cal L}_{\rm pNRQCD} &=& \int d^3r \;  {\rm Tr} \, \Bigg\{ 
{\rm S}^\dagger 
\left( i\partial_0 + \frac{\bfnabla^2_R}{4 m} +\frac{\bfnabla_r^2}{m}  - V_S
\right) {\rm S} \Bigg \},
\label{pNRQCD:Lag}
\eea 
${\rm S}$ is a non-local field, function of ${\bf r}$, ${\bf
  R}$ and $t$, $2\otimes 2$ in spin space and a $3 \otimes 3$ singlet in colour space.
The trace is taken over colour and spin indices.

All complications go into the potential $V_S$, which is a 
non-perturbative function of ${\bf r}$ to be determined by a non-perturbative 
matching procedure. In general $V_S$ contains also an imaginary part inherited 
from the matching coefficients of the 4-fermion operators of NRQCD. 
The matching condition reads
\bea
\langle \underbar{0}; {\bf r}' ,{\bf R}'|H_{\rm NRQCD}|\underbar{0}; {\bf r},{\bf R}\rangle
= \left(- \frac{\bfnabla^2_R}{4 m} - \frac{\bfnabla_r^2}{m}  + V_S\right)
\delta^{3}({\bf r}^\prime-{\bf r})\delta^{3}({\bf R}^\prime-{\bf R}).
\label{singlet}
\eea
The matching condition determines $V_S$ as a function of quantities defined in NRQCD
(the left-hand side of Eq.~(\ref{singlet})). Once $V_S$ has been determined,
one may calculate the solutions $\Phi_{H}({\bf r})$ and $E_H$ of the 
Schr\"odinger equation
\be
\left( - \frac{\bfnabla_r^2}{m} + V_S \right)\Phi_{H}({\bf r}) = E_H \Phi_{H}({\bf r}).
\label{schroedinger}
\ee

Using the optical theorem, the inclusive decay width to light particles (l.p.) is given by
\be
\Gamma_{H\to\rm l.p.} = - 2 \, {\rm Im}\, \langle H({\bf 0})| -{\cal L}_{\rm
  pNRQCD} | H({\bf 0}) \rangle.
\label{gammatot}
\ee
$| H({\bf 0}) \rangle$ stands for a quarkonium state in the rest-frame (${\bf P}$ = 0): 
\be
|H({\bf 0})\rangle = \int d^3r \int d^3R\; {\rm Tr}\, \left\{
\Phi_H({\bf r}){\rm S}^\dagger({\bf r},{\bf R})\right\}\ket{0},
\label{stateH0}
\ee
where $\ket{0}$ is the Fock subspace containing no heavy quarks but 
an arbitrary number of ultrasoft particles. 
Note that, since $\Phi_H$ and  ${\cal L}_{\rm pNRQCD}$ have been calculated 
through the matching procedure, Eq.~(\ref{gammatot}) provides, indeed, a practical 
tool to calculate the inclusive decay width. 
Explicit applications of Eq.~(\ref{gammatot}) have been worked out in 
\cite{Brambilla:2001xy,Brambilla:2002nu,Vairo:2002nh}.

\section{Radiative Transitions in pNRQCD}
Radiative transitions may be described in the same EFT framework that we have 
discussed so far by enlarging the gauge group to $SU_c(3)\times U_{\em}(1)$. 
This means that more degrees of freedom have to be taken 
into account (photons) and more operators added to the EFT Lagrangians. 
We will concentrate in the following on magnetic dipole transitions \cite{BJV05}.

At the level of the NRQCD Lagrangian, magnetic transitions are accounted for by replacing 
$i D_0 \to i D_0  - e e_Q A^\em_0$ 
and  $i {\bf D} \to i {\bf D} + e e_Q {\bf A}^\em$ in Eq.~(\ref{LNRQCD2f}), 
\bea
& & \hspace{-8mm}
{\cal L}_{2-f} \to  {\cal L}_{2-f} 
+ \frac{c_F^\em}{2 m} \psi^\dagger \bfsigma \cdot e e_Q {\bf B}^{\em} \psi
- \frac{2c_F^\em -1}{8 m^2} \psi^\dagger \bfsigma \cdot [-i{\bf D}\times, e e_Q {\bf E}^{\em}]\psi  
\nn \\ 
&& \hspace{-6mm}
+ \frac{c^\em_{W1}}{8 m^3} \psi^\dagger \{ {\bf D}^2, \bfsigma \cdot e e_Q
    {\bf B}^{\em} \}\psi
- \frac{ c_{W1}^\em -1}{4 m^3} \psi^\dagger 
 {\bf D}^i \bfsigma \cdot e e_Q {\bf B}^{\em} {\bf D}^i \psi
 \nn \\
&& \hspace{-6mm}
+\frac{c_F^\em -1}{ 8 m^3} 
\psi^\dagger \left( \bfsigma\cdot{\bf D} \; e e_Q {\bf B}^\em \cdot {\bf D}
+ {\bf D} \cdot e e_Q {\bf B}^\em \;  \bfsigma\cdot{\bf D} \right) \psi 
+ \dots 
+ [\psi \rightarrow i \sigma ^2 \chi^*],
\label{LNRQCD2fgamma}
\eea
and $\displaystyle {\cal L}_{\rm light} \to {\cal L}_{\rm light} 
- \frac{1}{4}F^{\mu\nu\,\em}F_{\mu\nu\,\em}$, 
where the gauge fields with upperscript ``$\em$'' are electromagnetic fields
and $e e_Q$ stands for the charge of the quark of flavour $Q$.

The coefficients $c_F^\em$ and $c^\em_{W1}$ are new matching coefficients
of the EFT associated with the electromagnetic couplings. 
Again, we have made use of reparameterization invariance  
to reduce their number. All coefficients are known at least at one-loop level
\cite{Manohar:1997qy}. 
In particular, we have\footnote{The coefficients get also QED
  corrections, but they are numerically negligible.} 
\bea
c_F^\em \equiv 1+ \kappa_Q^\em &=& 1+ \frac{4}{3} \frac{\als}{2\pi} + {\cal O}(\als^2)\,,
\label{kQ} \\
c_{W1}^\em &=& 1+ \frac{4}{3} \frac{\als}{\pi} 
\left(\frac{1}{12} + \frac{4}{3} \ln \frac{m}{\mu} \right)+ {\cal O}(\als^2)\,,
\eea
$\kappa_Q^\em$ is usually identified with the anomalous magnetic moment of the heavy quark. 

At the level of pNRQCD, magnetic transitions involving ultrasoft photons are described by adding to the Lagrangian
(\ref{pNRQCD:Lag}) the electromagnetic Lagrangian $- F^{\mu\nu\,\em}$ $\times F_{\mu\nu\,\em}/4$ and 
a term ${\cal L}_{\gamma\, \rm pNRQCD}$ responsible for the coupling of the quarkonium to the electromagnetic field:
\bea
&&\hspace{-8mm}
{\cal L}_{\gamma\, \rm pNRQCD} 
=  \int d^3 r \;  {\rm Tr} \, \Bigg\{ 
V_A^\em \; {\rm S}^\dagger {\bf r}\cdot e e_Q {\bf E}^{\em} {\rm S} 
\nn \\
&&\hspace{-6.5mm}
+ \frac{V^{\frac{\sigma\cdot B}{m}}_S 
}{2 m}
\left\{{\rm S}^\dagger , \bfsigma \cdot e e_Q {\bf B}^{\em}\right\} {\rm S} 
+ \frac{V^{(r\cdot \nabla)^2 \frac{\sigma\cdot B}{m}}_S
}{16 m}
\left\{ {\rm S}^\dagger , {\bf r}^i {\bf r}^j 
    (\bfnabla^i_R \bfnabla^j_R \bfsigma \cdot e e_Q {\bf B}^{\em})\right\} {\rm S} 
\nn \\
&&\hspace{-6.5mm}
+ \frac{V^{\frac{\sigma\cdot (r \times r \times B)}{m^2}}_{S}}{4 m^2\,r} 
\; \left\{{\rm S}^\dagger , \bfsigma\cdot\left[ \hat{\bf r} \times  \left(
  \hat{\bf r}\times e e_Q {\bf B}^{\em} \right) \right] \right\} {\rm S} 
+ \frac{V^{\frac{\sigma\cdot B}{m^2}}_{S}}{4 m^2\, r}
\; \left\{{\rm S}^\dagger , \bfsigma \cdot e e_Q {\bf B}^{\em}\right\} {\rm S} 
\nn \\
&&\hspace{-6.5mm} 
- \frac{
V^{\frac{\sigma\cdot \nabla \times E}{m^2}}_S}{16 m^2}  
\left[{\rm S}^\dagger,  \bfsigma \cdot \left[-i\bfnabla_R \times, e e_Q {\bf E}^\em \right]\!\right] {\rm S}
- \frac{V^{\frac{\sigma\cdot \nabla_r \times r\cdot \nabla E}{m^2}}_S
}{16 m^2} 
\left[ {\rm S}^\dagger,  \bfsigma \cdot \left[-i\bfnabla_r \times, 
{\bf r}^i (\bfnabla^i_R e e_Q  {\bf E}^\em) \right]\! \right] {\rm S}
\nn \\
&&\hspace{-6.5mm}
+ \frac{V^{\frac{\nabla_r^2 \, \sigma\cdot B}{m^3}}_S
}{4 m^3}  
\left\{ {\rm S}^\dagger , \bfsigma \cdot e e_Q {\bf
  B}^{\em} \right\} \bfnabla_r^2 {\rm S} 
+ \frac{V^{\frac{(\nabla_r\cdot\sigma)\, (\nabla_r\cdot B)}{m^3}}_S
1}{4 m^3} 
\left\{ {\rm S}^\dagger , \bfsigma^i \, e e_Q {\bf
  B}^{\em\,j} \right\} \bfnabla_r^i\bfnabla_r^j {\rm S} 
\nn \\
&&\hspace{-6.5mm} 
+ \frac{V^{\frac{\sigma\cdot (r \times r \times B)}{m^3}}_{S}}{4 m^3\, r^2} 
\left\{{\rm S}^\dagger , \bfsigma\cdot\left[ \hat{\bf r} \times  \left(
  \hat{\bf r}\times e e_Q {\bf B}^{\em} \right) \right] \right\} {\rm S} 
+ \frac{V^{\frac{\sigma\cdot B}{m^3}}_{S}}{4 m^3\, r^2}
\left\{{\rm S}^\dagger , \bfsigma \cdot e e_Q {\bf B}^{\em}\right\} {\rm S}
+ \cdots \Bigg\}\,.
\label{gammapNRQCD:Lag}
\eea
All gauge fields are calculated in the centre-of-mass coordinate ${\bf R}$.
The field ${\rm S}$ is understood as a singlet also under $U_{\em}(1)$ gauge transformations.

In the centre-of-mass of the initial quarkonium state, 
the power counting goes as follows: $\bfnabla_r \sim mv$, $r\sim 1/mv$, 
the electromagnetic fields associated to the external 
photons go like ${\bf E}^\em, {\bf B}^\em \sim k_\gamma^2$. The centre-of-mass derivative 
$\bfnabla$ acting on the recoiling final quarkonium state or emitted photon 
is of order $k_\gamma$, where $k_\gamma$ is the energy and momentum of the emitted photon.

The coefficients $V$ in Eq.~(\ref{gammapNRQCD:Lag}) are the matching
coefficients of pNRQCD. They encode high-energy contributions to the
electromagnetic couplings and are of the same nature as $V_S$ in 
Eq.~(\ref{pNRQCD:Lag}). In the strong-coupling regime they are determined 
by non-perturbative matching of 5-points Green functions involving 
two external  quarks, two external antiquarks and an external photon.
Let us consider the matching condition for the $1/m$ operators, it reads 
\bea
&&\hspace{-12mm}
\langle \underbar{0}; {\bf r}' ,{\bf R}'|\otimes \langle \gamma |
\left(
\frac{c_F^\em}{2 m} 
\int d^3x \, \psi^\dagger \bfsigma \cdot e e_Q {\bf B}^{\em} \psi
+ [\psi \rightarrow i \sigma ^2 \chi^*]\right) |0\rangle \otimes  |\underbar{0}; {\bf r},{\bf R}\rangle
= 
\nn\\
&& \hspace{-11mm}
\left(\!
\frac{V^{\frac{\sigma\cdot B}{m}}_S}{2m} + 
\frac{V^{(r\cdot \nabla)^2 \frac{\sigma\cdot B}{m}}_S}{16m}
\left({\bf r}\cdot\bfnabla_R\right)^2 
\!\right)\!\!
\left(\bfsigma^{(1)} + \bfsigma^{(2)}\right) \cdot 
\langle \gamma| e e_Q {\bf B}^{\em}|0\rangle 
\delta^{3}({\bf r}^\prime-{\bf r})\delta^{3}({\bf R}^\prime-{\bf R}).
\label{SBmatch}
\eea
Since corrections to the state $|\underbar{0}; {\bf r},{\bf R}\rangle$
involving derivatives or spins are $1/m$ suppressed (see
Eq.~(\ref{LNRQCD2f})), $\bfsigma \cdot e e_Q {\bf B}^{\em}$ 
effectively behaves as the identity operator. As a consequence, the 
electromagnetic matrix element decouples in the left-hand side. 
From the normalization condition (\ref{normstate}) it follows that 
\be
V^{\frac{\sigma\cdot B}{m}}_S = V^{(r\cdot \nabla)^2 \frac{\sigma\cdot
    B}{m}}_S  = c_F^\em.
\label{m11}
\ee
This is a rather remarkable result that holds to all orders in the strong-coupling
constant and non-perturbatively. It excludes that the $1/m$ magnetic coupling 
of the quarkonium field is affected by any soft contribution.
A fortiori, it excludes large anomalous non-perturbative corrections 
to this coupling. Similar arguments lead to the following exact results 
at order $1/m^2$:
\be
V^{\frac{\sigma\cdot (r \times r \times B)}{m^2}}_{S} = \frac{r^2}{2} V_S^{(0)\,\prime}, 
\quad 
V^{\frac{\sigma\cdot B}{m^2}}_{S} = 0,
\quad 
V^{\frac{\sigma\cdot \nabla \times E}{m^2}}_S = 
V^{\frac{\sigma\cdot \nabla_r \times r\cdot \nabla E}{m^2}}_S = 2c_F^\em -1,
\label{m21}
\ee
where $V_S^{(0)}$ is the static part of the $V_S$ potential. 
The first equality follows from the fact that Poincar\'e invariance 
protects the spin-orbit coupling \cite{Brambilla:2003nt,Brambilla:2001xk}.
The second one remarkably states that to all orders in the strong-coupling constant and non-perturbatively
the existence of an effective scalar interaction, which has been often advocated in
phenomenological models, is excluded. The third one that those matching
coefficients, like the one in Eq.~(\ref{m11}), get only hard contributions.

The matching of the $1/m^3$ terms is more complicated. One reason is that at this
order kinetic energy  and spin-dependent corrections affect 
the state $|\underbar{0}; {\bf r},{\bf R}\rangle$ and $\bfsigma \cdot e e_Q {\bf B}^{\em}$
does not behave anymore like the identity operator.

Once the matching has been completed, the transition width is given by:
\be
\Gamma_{H \to H^\prime \gamma} = \int \frac{d^3P^\prime}{(2\pi)^3} \frac{d^3k}{(2\pi)^3}\,
(2\pi)^4\delta^4(P_H-k-P^\prime)\, 
\overline{\big| {\cal A}\left[H({\bf 0}) \to H^\prime({\bf P}^\prime) 
\gamma({\bf k})\right] \big|^2},
\label{transitionwidth}
\ee
where 
\bea
{\cal A}\left[H({\bf 0}) \to H^\prime(-{\bf k}) \gamma({\bf k})\right] 
\; \delta^3({\bf P}^\prime + {\bf k})
= \bra{ H^\prime({\bf P}^\prime) \,\gamma({\bf k})}
- \int d^3R\, {\cal L}_{\gamma\,\rm pNRQCD}\ket{H({\bf 0})} 
\label{transitionamplitude}
\,.
\eea
The overline stands for the sum over the final-state polarizations and the
average over the initial state ones. $P_H = (M_H,{\bf 0})$ stands for the
four-momentum of the initial-state quarkonium of mass $M_H$.
The state $\ket{H({\bf P})}$ is the state (\ref{stateH0}) boosted by $-{\bf
  P}/M_H$. The Lorentz-boost transformations may be read from \cite{Brambilla:2003nt,Brambilla:2001xk}.

\section{Conclusions}
We have discussed in an unified framework inclusive and electromagnetic decays, and 
radiative transitions of heavy quarkonium in a regime where the typical momentum 
transfer is of order $\lQ$. Noteworthy, also in this situation 
suitable effective field theories may be constructed, systematic expansions
exploited and exact results derived. 
 
It seems rather unlikely that the non-perturbative matching, once completed at order $1/m^3$,
will support the formulas traditionally and universally used so far to describe radiative transitions 
at relative order $v^2$ and derived from phenomenological assumptions 
\cite{Grotch:1982bi,Grotch:1984gf}. This may possibly shade some light, for
instance, on the radiative transition data for the $\Upsilon$ system 
recently collected at CLEO \cite{Artuso:2004fp}, whose understanding is 
problematic in many phenomenological models.

\begin{theacknowledgments}
The author acknowledges the financial support obtained inside the Italian 
MIUR program  ``incentivazione alla mobilit\`a di studiosi stranieri e 
italiani residenti all'estero'' and the Marie Curie Reintegration Grant contract MERG-CT-2004-510967.
\end{theacknowledgments}

\end{document}